\documentclass[twocolumn,letterpaper,aps,prd,longbibliography,superscriptaddress,nofootinbib,floatfix,groupedaddress]{revtex4-1}
\usepackage{graphicx}	
\usepackage{color}
\usepackage{hyperref}

\usepackage{amsmath}
\usepackage{xspace}	

\newcommand{\pt}{\mbox{$p_T$}\xspace}

\newcommand{\sqs}{\mbox{$\sqrt{s}$}\xspace}
\newcommand{\pp}{\mbox{$p$+$p$}\xspace}
\newcommand{\pout}{\mbox{$p_{\rm out}$}\xspace}
\newcommand{\ptassoc}{\mbox{$p_{\rm T}^{\rm assoc}$}\xspace}
\newcommand{\dphi}{\mbox{$\Delta\phi$}\xspace}
\newcommand{\rmspout}{\mbox{$\sqrt{\langle p_{\rm out}^{2}\rangle}$}\xspace}
\newcommand{\pttrig}{\mbox{$p_{\rm T}^{\rm trig}$}\xspace}
\newcommand{\pion}{\mbox{$\pi^{0}$}\xspace}
\newcommand{\h}{\mbox{\text{h}$^\pm$}\xspace}

\begin{document}


\title{Nonperturbative Transverse Momentum Effects in Dihadron and Direct Photon-Hadron Angular Correlations}

\author{J.D. Osborn for the PHENIX Collaboration}
\affiliation{Department of Physics, University of Michigan, Ann Arbor, Michigan 48109}

\begin{abstract}
Two-particle angular correlations have long been used as an observable for measuring the initial-state partonic transverse momentum $k_T$. Sensitivity to this small transverse momentum scale allows nonperturbative transverse momentum dependent effects to be probed in high $p_T$ dihadron and direct photon-hadron correlations. The observable $p_{out}$, the out-of-plane transverse momentum component from a near-side $\pion$ or direct photon, is sensitive to initial-state $k_T$ and final-state fragmentation transverse momentum $j_T$ and thus can probe nonperturbative transverse-momentum-dependent effects. In the transverse-momentum-dependent framework, nearly back-to-back particle production in $p$+$p$ collisions with a measured final-state hadron has been predicted to break factorization due to the possibility of gluon exchanges with colored remnants in the initial and final states.  For this reason, the interacting partons are predicted to be correlated; however, there is so far no quantitative prediction for the magnitude of such effects. In this talk, recent measurements of dihadron and direct-photon hadron correlations in $p$+$p$ collisions at $\sqrt{s}$=510 GeV at the PHENIX experiment will be presented.

\end{abstract}

\maketitle



\section{\label{Intro}Introduction}

In quantum chromodynamics (QCD), parton distribution and fragmentation functions (PDFs and FFs) are used to describe the long distance bound state nature of hadrons. Historically PDFs and FFs have been taken to only be dependent on the longitudinal momentum fraction $x$ and $z$, respectively. In the last several decades this has been extended to include transverse momentum dependence so that the ``unintegrated" transverse-momentum-dependent (TMD) PDFs and FFs are written explicitly as dependent on both the longitudinal momentum and transverse momentum of the partons. Since the framework explicitly includes small transverse momenta, the reevaluation of important principles of QCD like factorization and universality has been necessary. \par

In particular, the role of color interactions due to soft gluon exchanges between participants of the hard scattering and remnants of the interaction in collisions involving a hadron have been found to have profound effects regarding these principles. For example, the Sivers function, a particular TMD PDF that correlates the transverse spin of the proton with the orbital angular momentum of the parton, was predicted to be the same magnitude but opposite in sign when measured in semi-inclusive deep-inelastic scattering (SIDIS) and Drell-Yan (DY)~\cite{collins_signchange}. This prediction arises from the different color flows that are possible between these two interactions due to the possibility of soft gluon exchanges in the initial-state and final-state for DY and SIDIS, respectively. Factorization of the nonperturbative functions is still predicted to hold in both SIDIS and DY.\par

In $p$$+$$p$ collisions where two nearly back-to-back hadrons are measured, soft gluon exchanges are possible in both the initial and final state since there is color present in both the initial and final states. In this process, factorization breaking was predicted in a TMD framework~\cite{trog_fact,collins_qiu_fact}. For processes where factorization breaking is predicted, individual TMD PDFs and TMD FFs become correlated with one another and cannot be written as a convolution of nonperturbative functions. The ideas behind the predicted sign change of certain TMD PDFs and factorization breaking result from the same physical process of soft gluons being exchanged between participants of the hard scattering and remnants of the collision. These predictions represent major qualitative departures from purely perturbative approaches which do not consider the remnants of the collision at all. \par

In calculations of TMD processes where factorization is predicted to hold, the Collins-Soper (CS) equation is known to govern the evolution of nonperturbative functions in a TMD framework with the hard scale of the interaction $Q^2$~\cite{cs1,cs2}. Contrary to the purely perturbative collinear DGLAP evolution equations, the CS evolution equation for TMD processes involves the Collins-Soper-Sterman (CSS) soft factor~\cite{css_soft}, which contains nonperturbative contributions. The theoretical expectation from CSS evolution is that momentum widths sensitive to nonperturbative transverse momentum should increase with the hard scale of the interaction. This can intuitively be thought of as due to an increased phase space for hard gluon radiation, and thus a broader transverse momentum distribution. This behavior has been studied and confirmed in several phenomenlogical analyses of DY and Z boson data (see e.g.~\cite{dy1,dy2,sidisdy}) as well as phenemonological analyses of SIDIS data (see e.g.~\cite{sidisdy,sidis1,sidis2}). Since the CS evolution equation comes directly out of the derivation of TMD factorization~\cite{css_tmd}, it then follows that a promising avenue to potentially observe factorization breaking effects is to study possible deviations from CSS evolution in processes where factorization breaking is predicted such as dihadron and direct photon-hadron angular correlations in $p$$+$$p$ collisions. \par



\section{Dihadron and Direct Photon-Hadron Angular Correlations}

Dihadron and direct photon-hadron angular correlations are both predicted to be sensitive to factorization breaking effects because there are hadrons in both the initial and final states, thus the potential for soft gluon exchanges in both the initial and final states exists. Additionally these processes can be treated in a TMD framework when the two particles are nearly back-to-back and have large $p_T$; a hard scale is defined with the large $p_T$ of the particles and the process is also sensitive to the convolution of initial-state and final-state small transverse momentum scales $k_T$ and $j_T$. Here $k_T$ refers to the initial-state partonic transverse momentum due to the confined nature of the partons and soft or hard gluon radiation, and $j_T$ refers to the final-state fragmentation transverse momentum due to soft or hard gluon radiation in the hadronization process. \par

 \begin{figure}[thb]
 \includegraphics[width=1.0\linewidth]{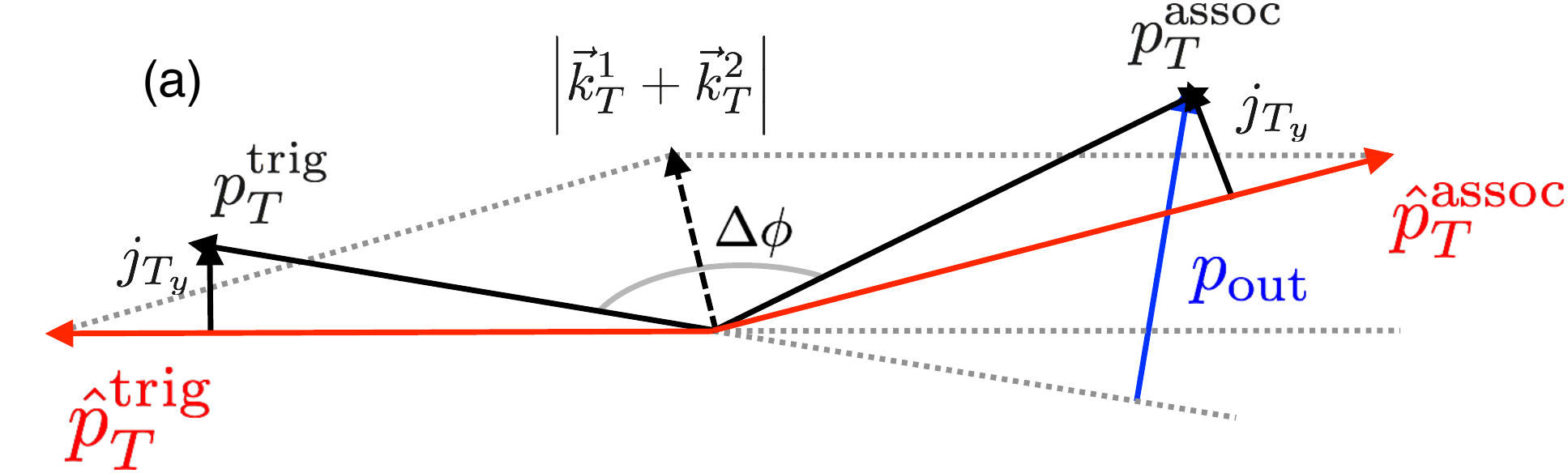} \\
 \includegraphics[width=1.0\linewidth]{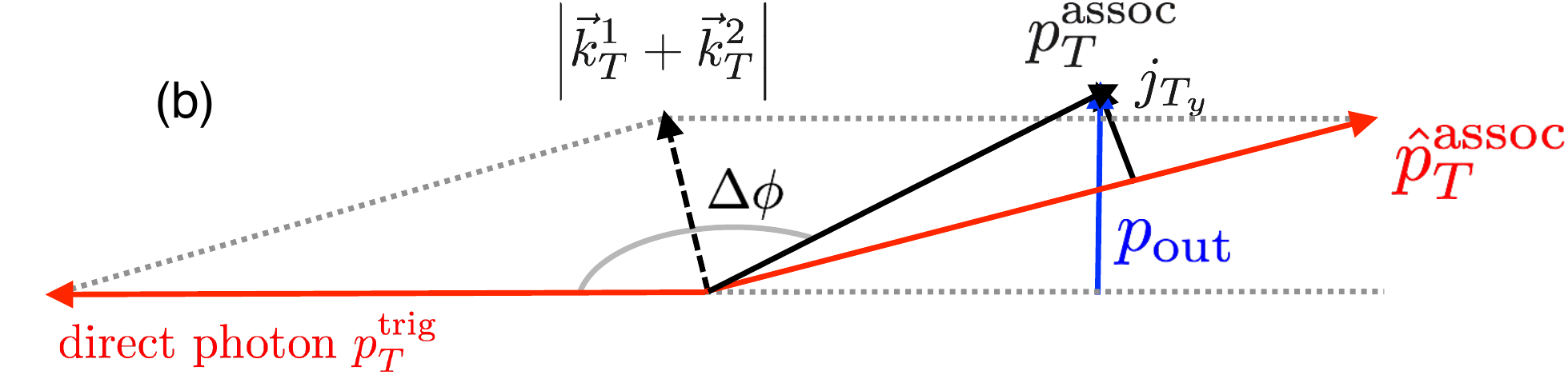}
 \caption{\label{fig:ktkinematics} 
 A schematic diagram showing event hard-scattering kinematics for (a) dihadron and (b) direct photon-hadron processes in the transverse plane. Two hard scattered partons with transverse momenta $\hat{p}_T^{\rm trig}$ and $\hat{p}_T^{\rm assoc}$ are acoplanar due to the initial-state partonic $k_T$, given by $|\vec{k}_T^1+\vec{k}_T^2|$. The scattered partons result in two fragmented hadrons that are additionally acoplanar due to the final-state transverse momentum perpendicular to the jet axis $j_{T_y}$. For (b) direct photon-hadrons only one away-side jet fragment is produced since the direct photon is colorless. The transverse momentum component perpendicular to the trigger particle's axis is labeled as $\pout$. 
 }
 \end{figure}

Figure~\ref{fig:ktkinematics} shows a kinematic diagram in the transverse plane for both dihadron (top) and direct photon-hadron (bottom) events. At leading order the hard scattered partons are exactly back-to-back, but due to initial-state $k_T$ the partons are acoplanar by some amount $|\vec{k}_T^1+\vec{k}_T^2|$. The fragmentation process introduces an additional transverse momentum component $j_{T_y}$ which is assumed to be Gaussian distributed about the parton axes such that $\sqrt{\langle j_T^2\rangle}=\sqrt{2\langle j_{T_y}^2\rangle}$. The transverse momentum component perpendicular to the trigger particle's axis is labeled as $\pout$ and is sensitive to initial and final state $k_T$ and $j_T$, where the trigger particle refers collectively to the direct photon or near-side hadron. Measuring the azimuthal angular separation between the near-side trigger particle and away-side associated particle allows calculating \pout with the following equation:
\begin{equation}
\pout = \ptassoc\sin\dphi
\end{equation}

The results presented were measured by the PHENIX collaboration at the Relativistic Heavy Ion Collider at Brookhaven National Lab. The PHENIX detector covers a pseudorapidity interval of $|\eta|<0.35$, and has two arms which in total span an azimuthal region of $\Delta\phi\sim\pi$. Lead scintillator and lead glass electromagnetic calorimeters provide measurements of isolated direct photons and neutral pions via their two photon decay. To measure away-side particles, the PHENIX detector employs a drift chamber and pad chamber tracking system that measures nonidentified charged hadrons. These results were measured from $p$$+$$p$ collisions at a center-of-mass energy of $\sqrt{s}=510$ GeV, and were recently submitted to the arXiv~\cite{ppg195}.

\section{Results}
\begin{figure*}[thb]
\includegraphics[width=1.0\linewidth]{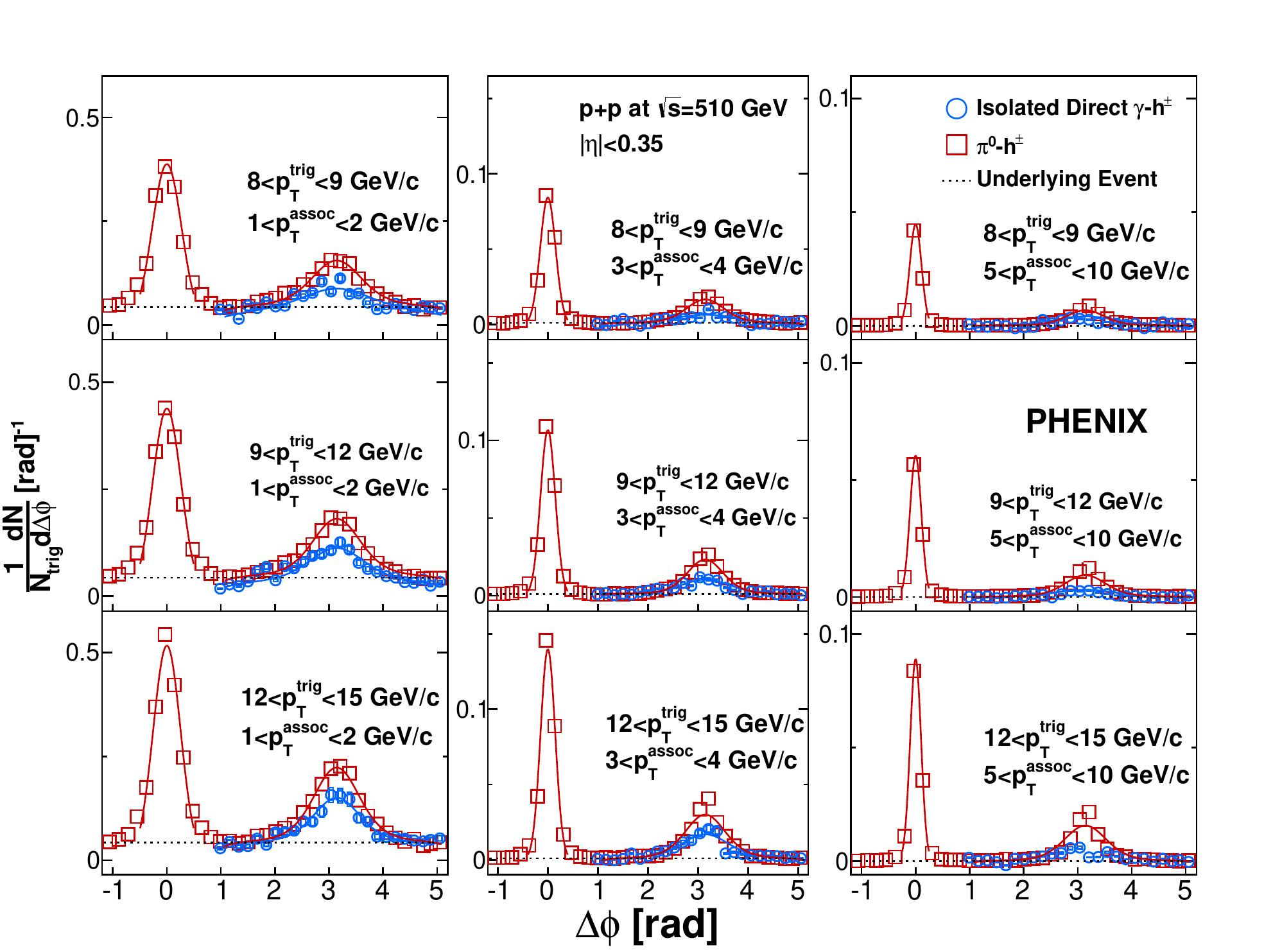}
\caption{\label{fig:dphis}
	Per trigger yields of charged hadrons as a function of $\dphi$ are shown in several \pttrig and \ptassoc bins for both \pion-hadron and direct photon-hadron correlations. The near side yield of the isolated direct photon-hadron correlations is not shown due to the presence of an isolation cut. The blue and red solid lines shown on the away-sides of the distributions are fits to extract the quantity \rmspout.  
	}
\end{figure*}

Figure~\ref{fig:dphis} shows the measured per-trigger yields as a function of \dphi for both \pion-hadron and direct photon-hadron correlations. The \pion-hadron correlations show the expected two jet structure, with two peaks that are nearly back-to-back around \dphi$\sim0$ and \dphi$\sim\pi$. The near side of the direct photon-hadron correlations is omitted due to the presence of an isolation cut on the direct photons; thus the near side is not physically interpretable. Additionally, the away-side jets are sensitive to the effects of $k_T$ broadening, so they are the yields of interest. The direct photon-hadron yields have a smaller yield than the \pion-hadron yields due to the direct photon emerging from the hard scattering at leading order; thus the direct photon-hadron correlations probe a smaller jet energy than the \pion-hadron correlations. \par

\begin{figure*}[thb]
\includegraphics[width=1.0\linewidth]{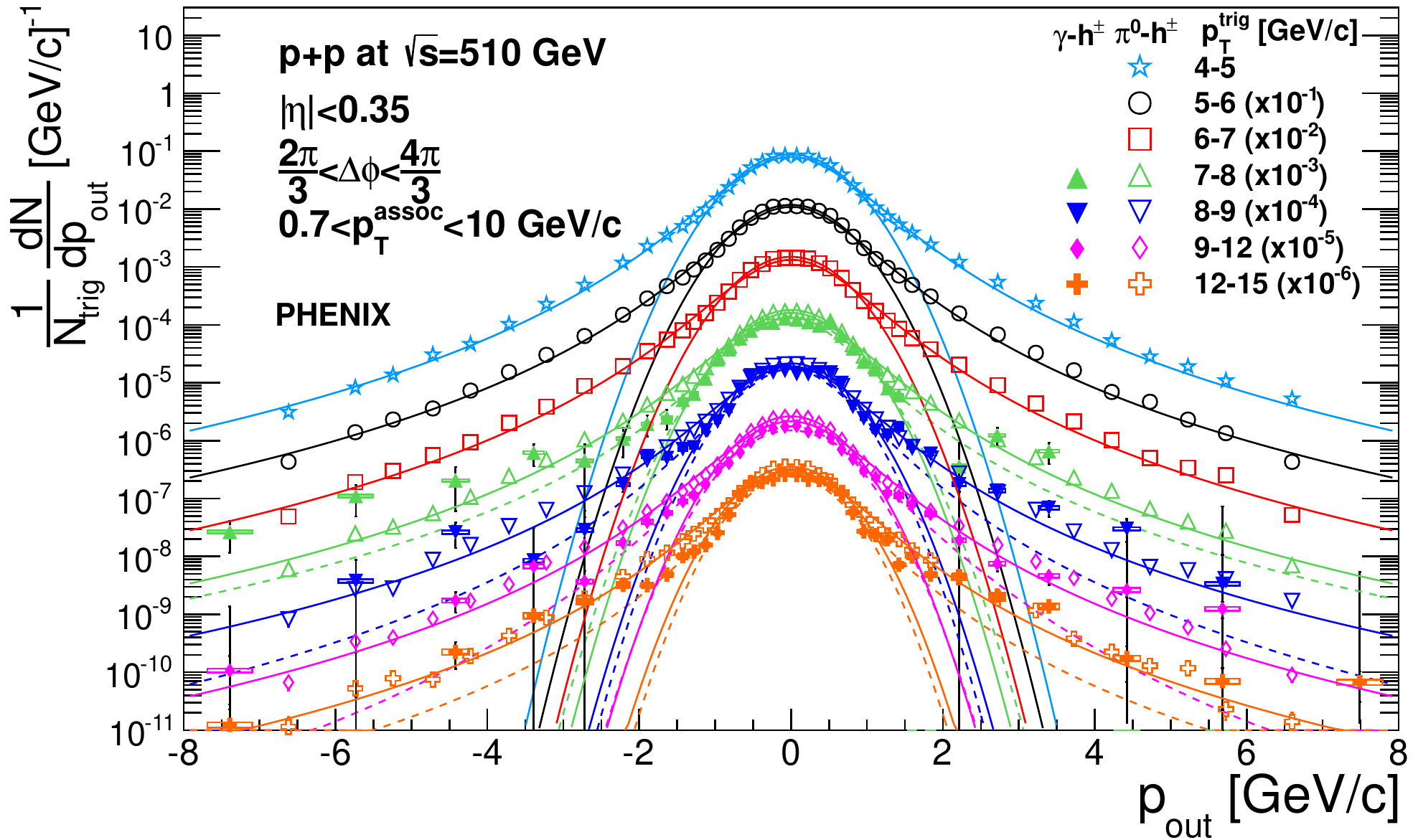}
\caption{\label{fig:pouts}
Per trigger yields of charged hadrons as a function of $\pout$ are shown in several $\pttrig$ bins for both $\pion$-$\h$ and direct photon-$\h$ correlations. The distributions are fit with a Gaussian function at small $\pout$ and a Kaplan function over the entire range. The Gaussian fit clearly fails after $\sim$1.3 GeV/c, indicating a transition from nonperturbatively generated $k_T$ and $j_T$ to perturbatively generated $k_T$ and $j_T$. 
}
\end{figure*}

The measured $\pout$ per-trigger yields are shown in Fig.~\ref{fig:pouts} for both $\pion$-$\h$ and direct photon-$\h$ angular correlations. The open points show the distributions for $\pion$-hadron correlations while the filled points show the distributions for the direct photon-hadron correlations. The distributions are constructed for only away-side charged hadrons that are sensitive to initial-state $k_T$ and final-state $j_T$. The $\pout$ distributions show two distinct regions; at small $\pout\sim0$ where the particles are nearly back-to-back a Gaussian shape can be seen, while at larger $\pout$ a power law shape is clear. These two shapes indicate a transition from nonperturbatively generated transverse momentum due to soft gluon emission in the Gaussian region to perturbatively generated transverse momentum due to hard gluon emission in the power law region. The distributions are fit with a Gaussian function at small $\pout$ and a Kaplan function over the entire distribution. The Gaussian function clearly fails at $\pout\sim$1.3 while the Kaplan function accurately describes the transition from Gaussian behavior to power law behavior. \par

\begin{figure}[thb]
\includegraphics[width=\linewidth]{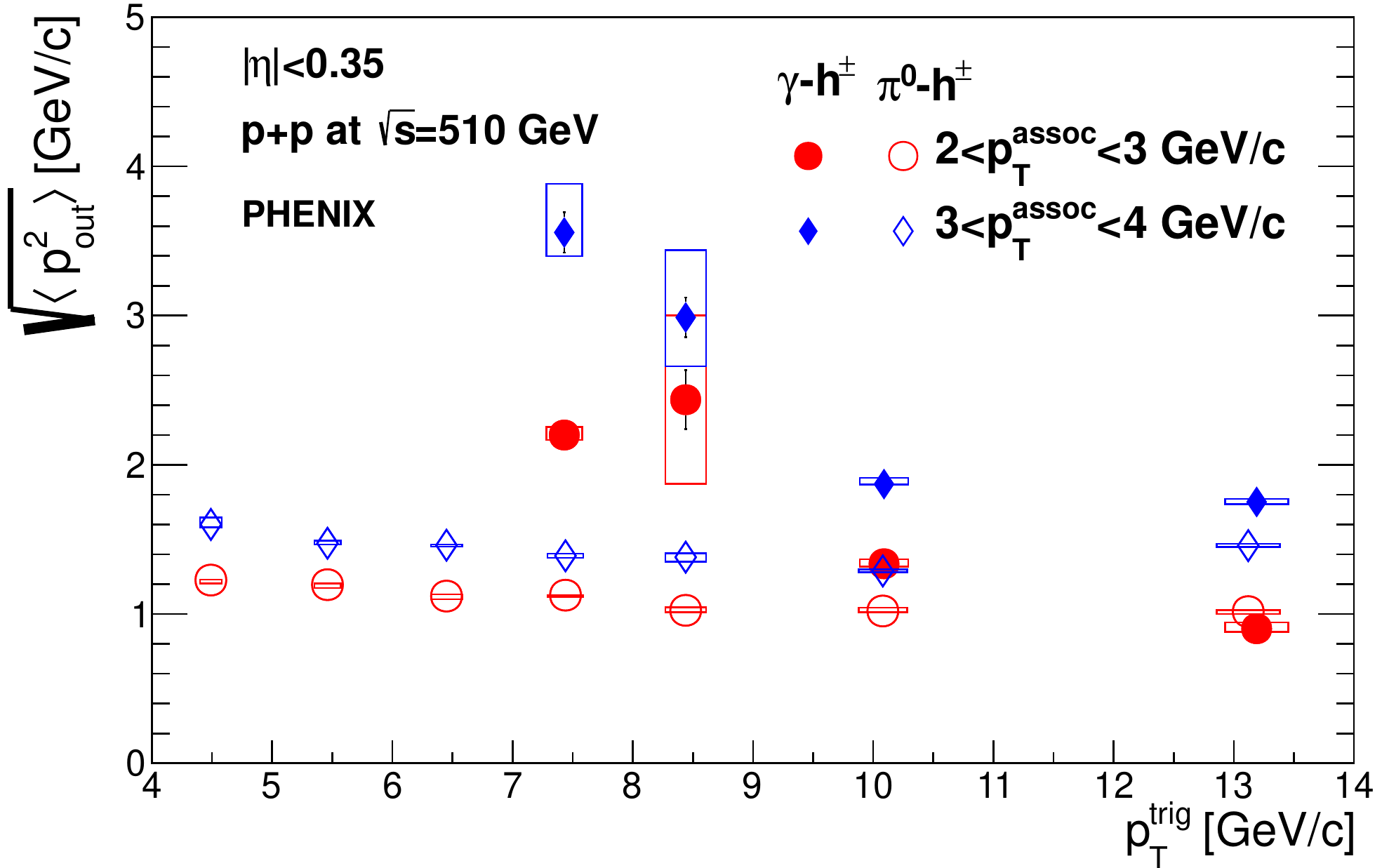}
\caption{\label{fig:rmspouts}
The extracted values of \rmspout are shown as a function of the interaction hard scale \pttrig for both \pion-hadron and direct photon-hadron correlations. The momentum widths decrease with the interaction hard scale, which is opposite the prediction from CSS evolution.
}
\end{figure}

To search for effects from factorization breaking, a comparison to the expectation from CSS evolution must be made using momentum widths which are sensitive to nonperturbative transverse momentum. To make a comparison, two different momentum widths were extracted from the measured correlations. Figure~\ref{fig:rmspouts} shows the root mean square of \pout as a function of $\pttrig$, which is extracted from fits to the entire away-side jet region in Fig.~\ref{fig:dphis}. The away-side fits can be seen in Fig.~\ref{fig:dphis} as red and blue solid lines around $\dphi\sim\pi$. The values of \rmspout clearly decrease for both the \pion-hadron correlations and direct photon-hadron correlations, which is the opposite of what is predicted from CSS evolution. The direct photon-hadron correlations show a stronger dependence than the \pion-hadron correlations in the same region of \pttrig. Since the values of \rmspout are extracted from the entire away-side jet region, \rmspout is sensitive to both perturbatively and nonperturbatively generated $k_T$ and $j_T$. While it is dominated by nonperturbatively generated transverse momentum since the majority of charged hadrons are in the nearly back-to-back region $\pout\sim0$ or $\dphi\sim\pi$, an observable that is sensitive to only nonperturbatively generated transverse momentum is the most ideal for comparisons to CSS evolution.\par

\begin{figure}[thb]
\includegraphics[width=\linewidth]{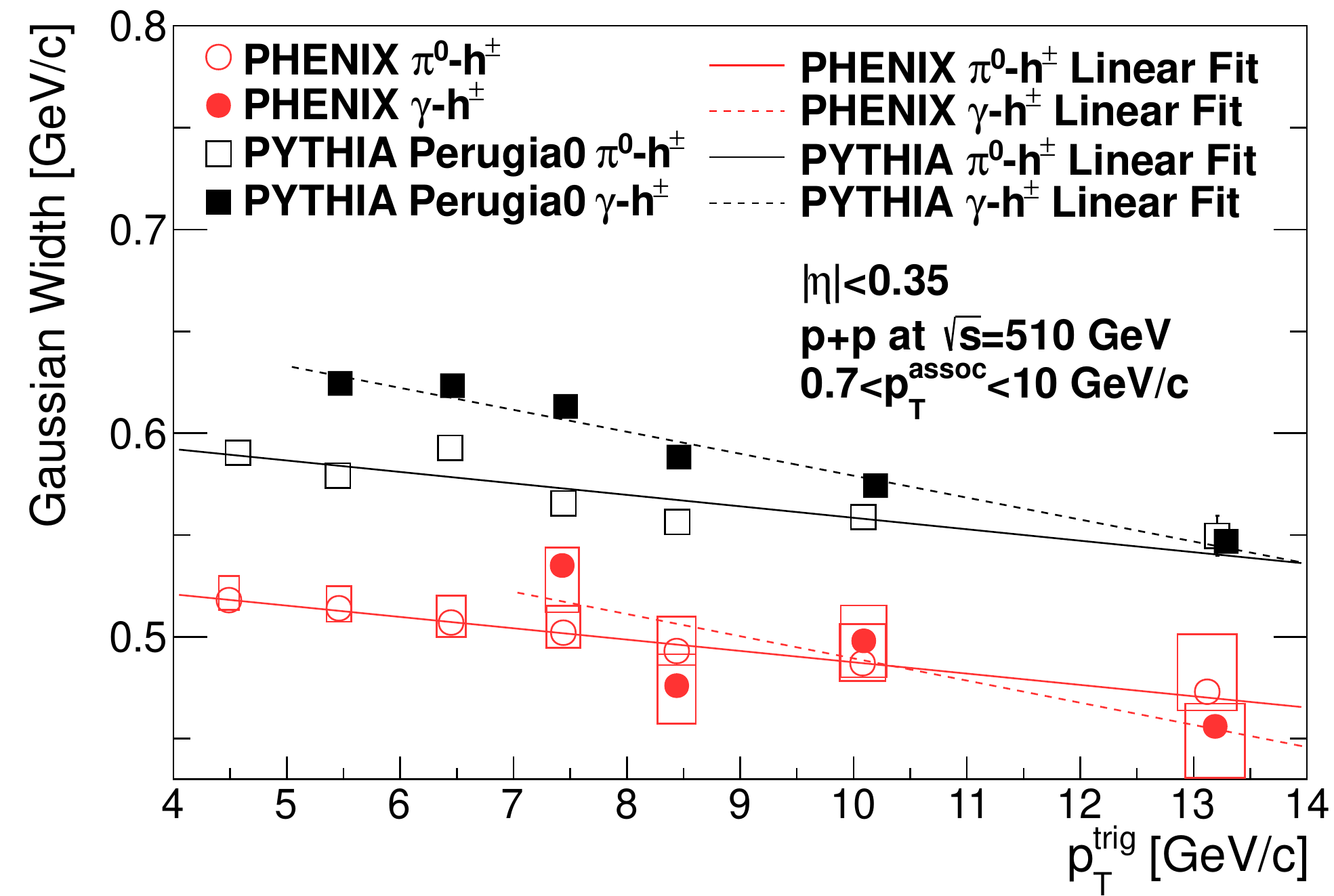}
\caption{\label{fig:widths}
The measured and {\sc pythia} simulated Gaussian widths from the \pout distributions are shown as a function of the hard scale \pttrig. The widths, sensitive to only nonperturbative transverse momentum, decrease with \pttrig. Surprisingly, {\sc pythia} nearly replicates the evolution behavior for both the direct photon-hadron and \pion-hadron correlations despite a $\sim$15\% difference in the magnitude of the {\sc pythia} widths.
}
\end{figure}

Since the Gaussian fits to the \pout distributions are taken in only the nearly back-to-back region, the widths of the Gaussian functions are momentum widths that are sensitive to only nonperturbative transverse momentum. Figure~\ref{fig:widths} shows the measured Gaussian widths as a function of the interaction hard scale \pttrig for both direct photon-hadron and \pion-hadron correlations. Similarly to the values of \rmspout from Fig.~\ref{fig:rmspouts}, the Gaussian widths decrease as a function of the hard scale, which is the opposite of the prediction from CSS evolution. The direct photon-hadron widths show a stronger dependence on \pttrig than the \pion-hadron widths, similar to the values of \rmspout. \par

Since theoretical calculations were not available, {\sc pythia}~\cite{pythia} was used to simulate direct photon-hadron and \pion-hadron correlations and study the evolution of momentum widths with the hard scale of the interaction. The Perugia0~\cite{perugia} tune was used since it was tuned to Tevatron Z boson data at small \pt; therefore it should adequately reproduce events with a small amount of total \pt. {\sc pythia} direct photon and dijet events were produced and the \pout distributions were constructed directly from the simulation in exactly the same way that they were measured in data. Although the magnitude of the widths from the simulation is roughly 15\% different in each bin, the simulation remarkably nearly reproduces the measured evolution of the Gaussian widths as seen in Fig.~\ref{fig:widths}. It is plausible that {\sc pythia} could be sensitive to the effects from factorization breaking due to the way a \pp event is processed. Unlike a standard perturbative QCD calculation, {\sc pythia} forces all particles to color neutralize in the event. This includes allowing initial and final-state gluon interactions, which are the necessary physical mechanism for factorization breaking and are additionally necessary to color neutralize all of the objects in the event. \par

\section{Conclusion}

In hadronic collisions where at least one final-state hadron is measured, factorization breaking has been predicted in a TMD framework. When color is present in both the initial and final states, soft gluon exchanges between participants in the hard scattering and the remnants of the collision are possible, leading to novel color flows throughout the entire scattering process. Nearly back-to-back dihadron and direct photon-hadron angular correlations in \pp collisions at \sqs=510 GeV from the PHENIX experiment at the Relativistic Heavy Ion Collider were measured to probe possible effects from factorization breaking~\cite{ppg195}. The transverse momentum component perpendicular to the near-side trigger particle, \pout, was used to compare predictions from CSS evolution. CSS evolution, which comes directly out of the derivation of TMD factorization~\cite{css_tmd}, predicts that momentum widths sensitive to nonperturbative transverse momentum should increase with increasing hard scale due to the broadened phase space for perturbatively generated gluon radiation. This dependence has been observed in phenomenological fits to both DY and SIDIS data~\cite{dy1,dy2,sidisdy,sidis1,sidis2}. \par

The measured correlations at PHENIX show the opposite dependence from the prediction of  CSS evolution; momentum widths in hadronic collisions where at least one final-state hadron is measured decrease with the interaction hard scale \pttrig~\cite{ppg195}. Remarkably, {\sc pythia} replicates this behavior in both direct photon-hadron correlations and \pion-hadron correlations. While {\sc pythia} certainly does not consider effects from factorization breaking as it relies on collinear factorization, the necessary physical mechanism that results in the predicted factorization breaking is present in {\sc pythia}; gluon exchanges with the remnants are possible in a {\sc pythia} simulated event since all colored objects are forced to color neutralize in any given event, unlike a standard perturbative QCD calculation. \par

\bibliography{SPIN_proceedings_bib}

\end{document}